\documentclass[smallextended]{svjour3}       % onecolumn (second format)
\smartqed 
\usepackage[T1]{fontenc}
\usepackage[bitstream-charter]{mathdesign}
\usepackage{amsmath}                                                                    % Math
\usepackage{xcolor}
\usepackage{graphicx}
\usepackage{commons}
\usepackage{upgreek, natbib}
\usepackage{xcolor}
\usepackage{lscape}
\usepackage{multirow}
\usepackage{bm}
\usepackage{booktabs}
\usepackage{tikz}
\usetikzlibrary{matrix,arrows,decorations.pathmorphing, arrows,chains,matrix,positioning,scopes, decorations.markings}
\usepackage{mathtools, tkz-euclide}
\usetkzobj{all}
\tikzset{join/.code=\tikzset{after node path={%
\ifx\tikzchainprevious\pgfutil@empty\else(\tikzchainprevious)%
edge[every join]#1(\tikzchaincurrent)\fi}}}
\tikzset{>=stealth',every on chain/.append style={join},
         every join/.style={->}}
\tikzstyle{labeled}=[execute at begin node=$\scriptstyle,
   execute at end node=$]
\title{A mixture model approach to infer land-use influence on point referenced water quality}
\author{Adrien Ickowicz \and 
Jessica Ford \and
 Keith Hayes 
}
\institute{CSIRO,
Hobart, 7004 TAS, Australia}

\begin{document}
\maketitle

%\printauthor
\begin{abstract}
The assessment of water quality across space and time is of considerable interest for both agricultural and public health reasons. The standard method to assess the water quality of a catchment, or a group of catchments, usually involves collecting point measurements of water quality and other additional information such as the date and time of measurements, rainfall amounts, the land-use and soil-type of the catchment and the elevation. Some of this auxiliary information will be point data, measured at the exact location, whereas other such as land-use will be areal data often in a compositional format. Two problems arise if analysts try to incorporate this information into a statistical model in order to predict (for example) the influence of land-use on water quality. First is the spatial change of support problem that arises when using areal data to predict outcomes at point locations. Secondly, the physical process driving water quality is not compositional, rather it is the observation process that provides compositional data. 
In this paper we present an approach that accounts for these two issues by using a latent variable to identify the land-use that most likely influences water quality. This latent variable is used in a spatial mixture model to help estimate the influence of land-use on water quality. We demonstrate the potential of this approach with data from a water quality research study in the Mount Lofty range, in South Australia.
\end{abstract}
\section{Introduction}
Accounting for spatial association is becoming an increasingly important topic in ecological data analysis. Spatially referenced data are typically observed either at points in space (point-referenced or simply point data) or over areal units such as counties or zip codes (block data). The change of support problem \citep[COSP, see][]{Gelfand2001} "is concerned with inference about the values of the variable at points or blocks different from those at which it has been observed". Change of support problems of interest include predicting the dependent variable for a particular area given values of explanatory variables measured at points at different locations within that area (see for example \cite{Zhu2003, Cressie2011}) or the reverse: predicting some point-estimation given area-wide observations of explanatory variables.
It is the later problem that usually occurs in the context of water quality monitoring \citep[see][]{Beck1987}. Typically the analyst is presented with point-referenced observations (of water quality parameters such as concentrations of metals or nutrients, that may be used to ensure compliance with potable water quality standards. A natural way to approach this problem statistically is to develop a model of the spatio-temporal dependencies, where the spatial dependence is captured by the covariance matrix of the error term, and the temporal dependence is captured either through seasonality parameters\citep[see][]{Lindstrom2011, Lindstrom2013} or via an auto-regressive process \citep[AR, see][]{Bakar2013}. 
Typically, however, other parameters such as land-use type will be known (or suspected) to have an effect on water quality, and should therefore be incorporated it into the model. At this point two complications may occur. The first arises because land-use information is often recorded as a compositional measure \citep[see][]{Aitchison2003, VandenBoogaart2013}. For example, land-use is often recorded as a vector showing the proportion of the catchment under each of the use-types. The second complication is the COSP. This arises because land use is measured over an area that includes the point measurements of water quality.. 
Various methods exists for predicting the effect of land use on water quality at catchment scales. One common method is the Soil and Water Assessment Tool \citep[SWAT, see][]{Arnold1995} is a sophisticated, continuously distributed simulation model. It operates on a daily time step and is designed to predict the effect of land use, land management practices, and climate change on the quality and quantity of surface and ground water (see \url{http://swat.tamu.edu/}). \\
SWAT assumes an in-depth knowledge of the Wthe mechanistic processes that govern water quality and quantity within a watershed. It requires the analyst to quantify the parameters that govern the rates of these processes (such as surface runoff, percolation, Evapotranspiration, etc.) This level of understanding and information, however, is not always available (often because the cost of acquiring it at large scales is prohibitive)). 
In view of this alternative statistical methods have also been developed, the simplest of which is referred as the 'lumped' approach \citep[see][]{Strayer2003, King2005}. The lumped approach treats compositional observations as covariates in a linear model, sometimes relying on transformations of the explanatory and/or response variables \citep[see][]{Buck2004} or the derivation of response indices or metrics \citep[see][]{Shen2014} to work within the confines of the linear modelling framework. The lumped approach also assumes  that "each portion of the catchment has equal influence" on the water quality \citep[][]{Peterson2011}. This approach, however, does not address the COSP and this may lead to incorrect estimates, particularly if the areas of the catchment partition are large. For example, if a land-use type such as "forested highlands" occupies a large proportion of a catchment, but never occurs close to a water quality monitoring site (often in high-order river stretches), then it would be reasonable to anticipate that it would have  only a small effect on the water quality at the monitoring site), whereas under the lumped approach its influence would be proportional to its area in the catchment.\\
One way to avoid this problem with the lumped approach is to incorporate a "distance to the water" measurement in the statistical model, and \cite{Peterson2011} presents a list of techniques to achieve this. The basic idea is to modify the effect of land-use area by incorporating a weight proportional to the inverse distance between the land-use type and the point of measurement. This weighting may also allow for additional considerations, for example it may incorporate the effects of flow accumulation and dilution effects through a river network\citep[][]{Hunsaker1995} or the outputs of simple hydrological models \citep[][]{Burcher2009}. These approaches require knowledge of the land-use distribution across the whole catchment, in order to build the distance matrix, and are sensitive to decisions about how to measure distance between areal units and points.\\
\cite{Tong2002} takes a different approach to this problem by first constructing a non-parametric test to exclude land-use types that are not correlated with the mean of the water quality parameters within a particular hydrological unit (but still ignoring the COSP), and then using the remaining land-use covariates in a process-based simulation model. This  has the advantage of reducing the model dimension, and also helps to solve the issue of non-influential land-use categories mentioned earlier.  When testing the correlation between land-use and water quality in the first stage, however, the spatial-temporal structure of the problem is ignored. This could cause misleading results For example, a land-use category may be uncorrelated with the mean of a water quality parameter, and be therefore discarded in the initial stage of the analysis, but it may have a seasonal influence that would be missed in the subsequent modelling. \\
%% Added a few sentences to clarify why this ATP krigging is still not addressing our concern.
Another technique specifically designed to deal with the COSP is area-to-point kriging, described in \cite{Kyriakidis2004} and \cite{Yoo2006}, and applied in \cite{Bonyah2013}. Area-to-point kriging is an interpolation technique,that provides an interpolated compositional value at each location of measurement. For a given location, each composition value is a weighted average of the surrounding lattices compositions. Each weight carries both the dimension of the lattice and its distance to the location of interest. While this limits the influence of a distant/small lattice composition on the location, the covariates still belong to the composition space when the response variable is not, by essence, subject to compositional covariates. 
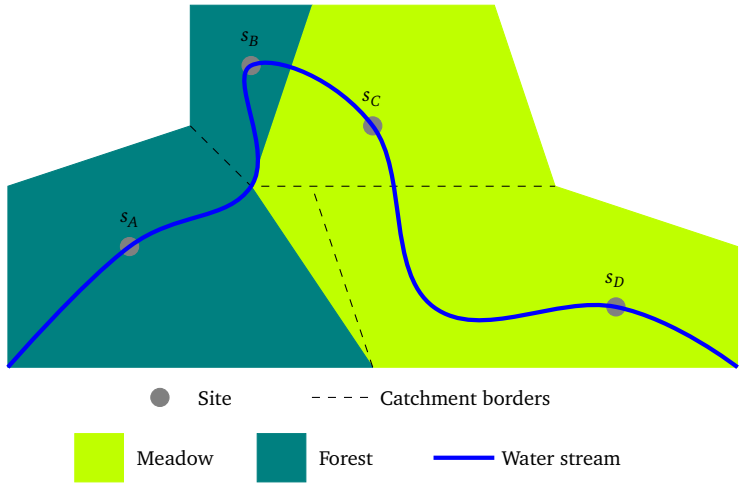
\begin{figure}[htb!]
\centering
\caption{Measurements in the same catchment are not subject to the same land-use influence. $s_B$ and $s_C$ are specific examples. }
\begin{tikzpicture}[scale = 0.8]

% LandUse

\draw[fill, teal] (-2,1) -- (-1,4) -- (-3,4) -- (-3,1);
\draw[fill, lime] (-2,1) -- (3,1) -- (2,4) -- (-1,4);
\draw[fill, lime] (0,-2) -- (-2,1) -- (3,1) -- (6,0) -- (6,-2) -- (0,-2);
\draw[fill, teal] (0,-2) -- (-6,-2) -- (-6,1) -- (-3,2) -- (-2,1) -- (0,-2);

% Subcatchments borders

\draw[thin, dashed]  (-3,2)--(-2,1)--(3,1);
\draw[thin, dashed]  (0,-2)--(-1,1);

% Nodes for stream

\node (s1) at (-2,1) {};
\node (s2) at (-6,-2) {};
\node (s3) at (1,-1) {};
\node (s4) at (6,-2) {};

% Measurement sites

\node[label={$s_A$}] (mA) at (-4,0) {$\bullet$};
\node[label={$s_B$}] (mB) at (-2,3) {$\bullet$};
\node[label={$s_C$}] (mC) at (0,2) {$\bullet$};
\node[label={$s_D$}] (mD) at (4,-1) {$\bullet$};
\draw[fill, gray] (mB) circle [radius = 0.15] ;;
\draw[fill, gray] (mC) circle [radius = 0.15];;
\draw[fill, gray] (mA) circle [radius = 0.15];;
\draw[fill, gray] (mD) circle [radius = 0.15];;

% Stream

\draw[blue, ultra thick]  plot[smooth, tension=.7] 
coordinates {(s2) (mA) (s1) (mB) (mC) (s3) (mD) (s4)};

%%% Legend %%%

\draw[fill, lime]  (-4.9,-3.1) rectangle (-4.1,-3.9);
\node at (-4.5,-3.5) {};
\node[right] at (-4,-3.5) {Meadow};

\draw[fill, teal]  (-1.9,-3.1) rectangle (-1.1,-3.9);
\node at (-1.5,-3.5) {};
\node[right] at (-1,-3.5) {Forest};

\draw[blue, ultra thick] (1,-3.5) -- (2,-3.5);
\node at (1.5,-3.5) {};
\node[right] at (2,-3.5) {Water stream};

\draw[black, thin, dashed] (-1,-2.5) -- (0,-2.5);
\node at (-0.5,-2.5) {};
\node[right] at (0,-2.5) {Catchment borders};

\node at (-3.5,-2.5) {$\bullet$};
\draw[fill, gray] (-3.5,-2.5) circle [radius = 0.15];;
\node [right] at (-3,-2.5) {Site};

\clip (-7,-5) rectangle (7,5);

%here_we_make_such_a_grid
%\draw[help lines,color=black!90](-7,-7)grid(7,7);

\end{tikzpicture}
\label{fig:schematic}
\end{figure}
In this paper we develop a mixture modelling approach to estimate the effect of land use on water quality, using the latent variable to identify which land-use type influences water quality at monitoring site. Mixture models have been studied for a long time, and their strengths and weaknesses are well known \citep[see][for a comprehensive review]{McLachlan2000}. Mixture models with compositional data, however, are not common, although some examples can be found within the existing literature \citep{Ongaro2008, Meinicke2011}. 
Mixture models offer some advantages in this context, for example, accounting for any spatial dependence between the latent variables in several ways, including:
\begin{itemize}
\item A mixture of expert with spatial random effects \citep{Neelon2014};
\item A mixture with a discrete or continuous (Markov random field - Potts or Gaussian process model) prior on the latent variable \citep{Woolrich2005}.
\end{itemize}
In this article, we demonstrate how a modified version of the latter approach provides an alternative to the methods described previously. The main advantage of our approach is that it deals with the COSP while also capturing the spatial and seasonal effects of land-use type on water quality. The disadvantages of this approach are that it limits the land-use influence to one type per site, for any given water quality parameter, and does not include an effect for the area of the land use type. 
 
The article is organized as follows. In section \ref{sec:model}, we present the model used to infer the spatio-temporal effects of land use types on water quality. In section \ref{sec:estimation} we present two estimation approaches. The first uses the Expectation-Maximisation \citep[EM, see][]{Dempster1977} algorithm. The second adopts a Bayesian hierarchical approach. In section \ref{sec:results} we compare the results obtained with our approach to the results obtained with an existing model drawn from the literature. We apply our method to a dataset from a water monitoring program from sub-catchment of the Mount Lofty Range in South Australia \citep[see][for details of the program]{Ford2015}. 
\section{Data and problem}
The Mount Lofty Ranges (MLR) are important in South Australia (SA) because they  provide significant water resources to a range of stakeholders, including agricultural landholders, secondary industries and potable water suppliers and consumers. A draft Water Allocation Plan (WAP) for the MLR was released in 2010-2011. In addition to the WAP the SA government identified the need for improved water quality in the MLR catchments through the Water Quality Improvement an the Water for Good programs \cite{Ford2015}. As part of these programs, water quality was monitored in $18$ sites over a period of $14$ years ($1998-2012$). Measurements were not collected every day due to resource limitations, but the frequency of observations was at least once per week. Various parameters including basic physico-chemical variables such as turbidity, dissolved oxygen, EC, pH and temperature as well as more investigation-specific parameters such as dissolved organic carbon (DOC) \cite{Varcoe2010} and nutrient and pesticide concentrations \cite{Cox2012} were recorded. These programs also recorded the land-use types (as a composition variable) in the catchments that contribute to the water being measured at each of the 18 monitoring sites. The land-use type  is  recorded within a hierarchical classification scheme. The first (coarsest) level of the hierarchy distinguishes 6 broad categories of land use. The second level splits this classification into 32 more detailed categories, and the third (finest) level further sub-divides the land use into 85 categories. In this analysis we use the 8 most well represented (and interpretable) land use categories from the second level primarily because the number of measurement sites was so small (to do otherwise would result in a very sparse design matrix and in over-parametrization issue).\\
Assessing the influence of land-use on total nitrogen with these data proves to be difficult for the following reasons:
\begin{enumerate}
\item We don't know what the land-use type is in the immediate vicinity of the measurement site - we only know the composition of land-use types in the catchment that the station is located in;
\item We don't know the distance along the stream network between measurement sites. We could calculate the Euclidean distance but this is not an optimal metric;
\end{enumerate}
Our analysis aims to identify the effect of land-use type on water quality, and thereby provide a mechanism to predict water quality  at sites in catchments that have not been monitored despite the aforementioned limitations. This analysis was designed to form part of a wider assessment of the risks of exceeding water quality parameters across the MLR, particularly under low flow conditions \cite{Ford2015}. Here we use the concentration of total nitrogen measured at $16$ sites from $2008$ to $2013$ to demonstrate application of the statistical model. 
\section{Modeling}
\label{sec:model}
\subsection{Framework}
Our data consists of geo-referenced, time stamped observations of Nitrogen concentration at 16 sites in the MLR. This immediately suggests the need for a spatio-temporal analysis. \cite{Szpiro2010} \citep[and then][]{Sampson2011, Lindstrom2011} propose the following general approach model for this type of data:
\begin{eqnarray}
\label{eq:spt_model}
\mathsf{y}(s,t) &=& \sum_j \mathsf{f}_j(t) {\bm   \beta}_j(s) + {\bm \nu}(s,t)
\end{eqnarray}
where ${\sf y}(s,t)$ denotes the observation at site $s$ and time $t$, $\beta_j \sim MVN(\tilde{\beta}_j, \Sigma_{\beta})$ captures the (spatially varying) effect of site-specific covariates  and ${\bm \nu} \sim N(0,\Sigma_{\nu})$ is a space-time residual field. The temporal variation in the data is decomposed into three basis functions $\mathsf{f}_j, j = 1 \cdots 3$ representing long term trend, seasonal effects and random variation respectively \citep[as detailed in][]{Cleveland1990} and $\tilde{\beta}_j$ captures the mean effect of site specific covariates on each of these components of temporal variation. 
In our analysis the relevant site-specific covariates are the land-use types in the catchment(s) that influence the water quality at a monitoring station.. In the general model, this 'land-level' information is represented by $Z_j$, such that: 
\begin{eqnarray}
\label{eq:latent}
{\bm \beta_j}(s) &=&  \alpha_j Z_j(s) + \eta(s)
\end{eqnarray}
where $\eta(s) \sim N(0, \Sigma_{\eta})$. It is clear from Equation~\ref{eq:latent} that :
\begin{itemize}
\item the temporal basis functions are the same at each location. The effect of the land-level covariates (e.g. land-use type) is modelled through $\alpha_j$ hence these covariates can only influence the amplitude of the basis function;
\item the spatial structure of the influence of the land-level covariates relies on the spatial correlation structure of $\beta_j$. Hence, it can be described using a stationary universal Kriging;
\item a common, and important simplification, stipulates that $\Sigma{\eta})$ is diagonal. This approach treats the water quality at each monitoring site as spatially independent, influenced only by (for example) the land-use type in its immediate catchment, and ignores the influence that other catchments might be have by virtue of the fact that monitoring sites may be connected by a stream network.
\end{itemize}
In this paper we present an approach that aims at improving this model in the following ways:
\begin{enumerate}
\item Land use types are typically recorded as a proportion of the surrounding catchment. This means that for each measurement site, a set of land-use types is observed, with the presumption that the probability the land-use types influences water quality is related in some way to the proportion of the catchment that they occupy. . In these circumstances, we believe that a mixture model is more appropriate than transforming the land-use observation using common transformations \citep[see][for a list of possible transformations.]{Aitchison2003}.
\item We relax the single temporal basis function assumption by allowing each mixture component to have its own set of temporal basis functions. For instance, given a land-use type $k$, measurements at site $i$ and time $t$ can be model using the following equation:
\begin{eqnarray}
y_i(t) | k &=& \sum_j \mathsf{f}_j(k,t) {\bm   \beta}_j(i) + {\bm \nu}
\end{eqnarray}
\item The spatial structure still relies on the land-use type, but we now use a neighbouring structure, of the type used in image analysis because we believe that modelling the spatial structure through the Euclidean distance between measurement sites is not optimal in our problem. Our interpretation is that the spatial continuity of land usage across multiple catchments is the key factor leading the Total nitrogen measurements characteristics. This translate into a neighbouring structure for the spatial model because of the lattice form of the catchments.
\end{enumerate}
\subsection{Incorporating land-use as a latent variable}
The general modelling framework we adopt is a point process model – i.e. it assumes that the land-use predictors in $\sf{Z}$ are observed at each of the locations $s_i$.. Land-use information, however, is most often defined over areas often with a significant size. This is a typical spatial misalignment problem, referred to as change of support problem in the introduction.\\
The land-use information collated in the MLR is presented as a compositional observations of land use types in the immediate catchment of the monitoring site, that is,
\begin{eqnarray}
\mathsf{z}_k(S_i) &=& \frac{1}{\vert S_i \vert }\int_{u \in S_i} \mathsf{z}_k(u) du 
\end{eqnarray}
where $\vert S_i \vert$ is the area for sub-catchment $i$, and$\mathsf{z}_k(u)$ an indicator function for the presence of land-use $k$ at the point location $u$ (in $S_i$). 
Ideally, we would build our model on the knowledge of $\mathsf{z}_k(s)$, where $s$ is the location of the monitoring station. For example, if we look at Figure \ref{fig:schematic}, $\mathsf{z}_F(s_A) = 1$, and $\mathsf{z}_F(S_A) = 1$ for land use type $k =$'Forest'. Which is good for the model, as this avoids the spatial misalignment issue. However, if we look at the second location, $\mathsf{z}_F(s_B) = 1$ for land use type 'Forest', but $\mathsf{z}_F(S_B)$ is only $0.3$ for this land use type. The majority of land-use in area $S_B$ is 'Meadow', and this could result in model estimates of $\mathsf{z}_F(s_B) = 0$, leading to inconsistency and bias in the resultant estimators. \\
What we propose is a mixture model approach. The rationale is the following. Given a single sample location, the land-use at this location has the biggest influence on the water quality measurements. This means that the measured water quality has a value "linked" to the land-use type. Then for multiple sites, the resulting measurements are sampled\footnote{\emph{sampled} is used here because we don't know the land-use type at the locations of measurement.} from a mixture distribution. The number of distributions in the mixture is then equal to the number of land-use types in the entire area. 
\subsection{Spatio-temporal structure of the model}
As stated in the previous sections, this problem is fully spatio-temporal and possible correlations have to be integrated in the model. We list below the solutions chosen to account for these correlations.
\subsubsection{Temporal dependency}
The model presented by \cite{Szpiro2010} supports the idea that the temporal variation remains globally identical over the spatial domain, making it possible to fit a smooth temporal function prior to fitting the full model. The spatio-temporal residuals $\nu$ are then assumed to be independent in both space and time. This assumption, however, is very strong. We therefore  develop an alternative modelling approach that allows for seasonality over the spatial domain, that can be affected by the land-use type at each measurement location, both in terms of amplitude, phase and shape. Note that the original model assumes that this influence affects amplitude alone. Our new model introduces temporal basis functions for each land-use type. These basis functions are calculated using a seasonal-trend (STL) decomposition approach introduced by \cite{Cleveland1990}. In this approach the time series of observations are decomposed into trend and seasonality components using a Loess smoother. In our model, the time series for each land-use is extracted from the time series of each site using the latent variable. Then, for a given land-use type, the STL decomposition is applied to each site time-series associated to that land-use through the latent variable.
\subsubsection{Spatial dependency}
It is possible to make multiple assumptions regarding the spatial structure. The most common approach is to assume that neighbouring sites are correlated with a correlation value that is a function of the Euclidean distance between sites. Recent works \citep{Peterson2007}  suggests that this model is not optimal when we dealing with data on a stream network , as the euclidean distance between sites is less relevant than the distance measured along the network, hence \cite{Peterson2007} suggest the use of stream distance rather than euclidean distance.
These two approaches limit the correlation structure to the measurements. We believe that some of the spatial dependency observed in the data come from the spatial structure of the land-uses. The solution presented in this paper is to model the spatial dependency through the latent variable of the mixture, making sure that neighbouring sites (in the sense of common border) are more likely to have identical land-uses. In the EM technique, this can be achieved by adding a penalization term. In the Gibbs sampling technique we model the latent variable of the mixture using a Potts model (also called 2-d Ising model). 
\subsection{Final model}
With the changes and assumptions described previously, the model equation becomes:
\begin{eqnarray}
\label{eq:spt_model_mixt}
\mathsf{y}(s,t) &=& \sum_j \mathsf{f}_j(z(s),t) {\bm   \beta}_j(s) + {\bm \nu}(s,t)
\end{eqnarray}
where the temporal basis function $\mathsf{f}$ is now land-use (location) dependent, and $\beta$ can be understood as a random effect parameter for each location-temporal basis pair. $z(s)$ is the latent indicator for which land-use is associated to site $s$.
\begin{figure}
\caption{Schematic figure of the model structure. The spatial correlation can be included at different levels of the model, leading to different likelihood and estimation tools. Model A include the spatial correlation at the baseline level ($y_0$), implying natural spatial correlation without the effect of land-use. This is the model most often used in the literature. Model B, which we use here, introduces the spatial correlation at the covariate level (land-use). Merging the two models can be considered, but one has to carefully monitor for over-fitting by doing so. }\label{fig:sch_model}
\centering
\vspace{1cm}
\begin{tikzpicture}[scale=.8]
        \node            (a) at (0,-5.5)  { Baseline WQ  };
        \node            (b) at (4,-5.5)  { Covariates };
        \node            (c) at (8,-5.5)  {Seasonality};
        \node            (e) at (12,-5.5)  { Error };
        
%       \node            (aa) at (0,-2.5)  { $y_0(s_i)$ };
        \node            (bb) at (4,-2.5)  { $ z(s_{i}) $ };
        \node            (cc) at (8,-2.5)  { $f_{j}(t)$ or $f_{j}(z(s_{i}),t)$};
        \node            (dd) at (12,-2.5)  { $\nu_i$};
%       \node            (ee) at (12,-1)  { $y(s_i,t)$};
        
        \node            (am) at (0,-1)  { $y_0(s_i)$ };
        \node            (bm) at (4,-1)  { $\beta_{j}(s_i)$ };
        \node            (cm) at (8,-1)  { $\sum f_{j}(z(s_i), t) \beta_{j}(s_i)$};
        \node            (dm) at (12,-1)  { $y(s_i,t)$};
        
        \node           (acov) at (0,-4)  { $\Sigma_\eta$ };
        \node            (bcov) at (4,-4)  { $\Sigma_Z$ };
        \node            (ecov) at (12,-4)  { $\Sigma_\nu$ };
        
        \draw[->] (am) edge (bm);
        \draw[->] (bm) edge (cm);
        \draw[->] (cm) edge (dm);
        
        \draw[->] (acov) edge (am);
        
        \draw[->] (bcov) edge (bb);
        \draw[->] (bb) edge (bm);
        
        \draw[->] (cc) edge (cm);
        
        \draw[->] (ecov) edge (dd);
        \draw[->] (dd) edge (dm);
        
\draw [rounded corners,blue, dashed, ultra thick] (-1,0) -- (-1,-5)-- (1,-5)-- (1,-2.8) -- (11,-2.8) -- (11,-4.5) -- (13,-4.5) -- (13, 0) -- cycle;
\draw [rounded corners,orange, dashed, ultra thick] (-0.9,-0.1) -- (-0.9,-2.9)-- (3,-2.9)-- (3,-5.1) -- (5, -5.1) -- (5, -2.9) -- (11.1,-2.9) -- (11.1,-4.6) -- (13.1,-4.6) -- (13.1, -0.1) -- cycle;   
        
        \node[text = blue]     (modA) at (0,-4.7)  { Model A };
        \node[text = orange]     (modB) at (4,-4.7)  { Model B };       
        
%\draw[->] (dm) edge (em);
        %       \draw[->] (e) edge (b);
        %       \draw[->] (f) edge (c);
        %\draw[->, dashed] (a) edge (c);
        %\draw[->, dashed] (e) edge (c);
        %\stoptikzpicture
        %\stoptext
        \end{tikzpicture}
\end{figure}
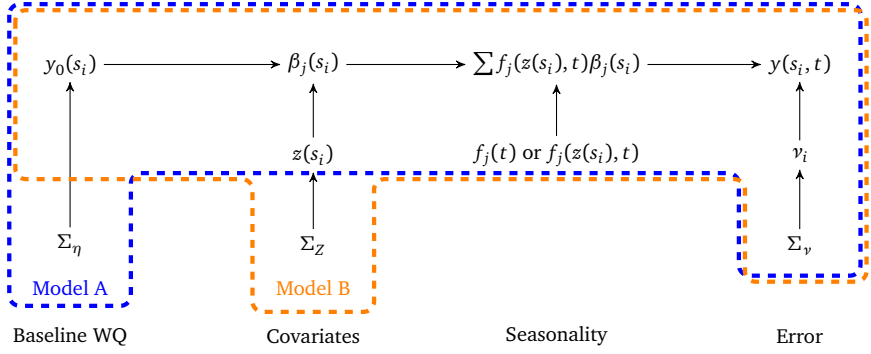
\section{Estimation}
\label{sec:estimation}
The model described in Eq. \ref{eq:spt_model} is very simple to express in the likelihood format, as all the components are Gaussian. This leads to a linear log-likelihood, for which maximization is performed quite straightforwardly \citep[see][for details about the maximization procedure, including some simplification tricks]{Lindstrom2013}.\\
The introduction of the mixture model on top of the spatial dependency (see Eq. \ref{eq:spt_model_mixt}) requires us to use an EM algorithm or a dedicated Bayesian hierarchical model. The following sections provide a description of the two possible estimation solutions.
\subsection{Pre-requisite}
There are two elements that are not estimated by the model, but need to be incorporated in order to perform the estimation: the temporal basis; and the neighbouring structure.
\paragraph{Temporal basis functions}
Preliminary temporal basis functions $\bar{f}_{ij}(t)$ are first estimated at each site using a Loess seasonal and trend decomposition \citep{Cleveland1990}. The basis function for each land-use is then calculated from these site specific decomposition using a linear equation,
\begin{eqnarray}
f_j(k, t) &=& (\sum_i 1_{(Z_i = k)} \bar{f}_{ij}) / \sum_i  1_{(Z_i = k)}
\end{eqnarray}
for function $j$, and land-use $k$. That equation also implies that in the estimation procedure, each time the latent variables are updated,  the temporal basis functions are also updated.
\paragraph{Neighbouring structure}
In order to compute the spatial dependency of the data, we need to identify the spatial structure. In this paper, we assume that the latent variables are carrying all the spatial dependency, by virtue of their neighbourhood relationship. This structure is achieved through the construction of a Voronoi lattice, where the cell centers are the monitoring sites.
\subsection{The EM approach}
The EM algorithm was initially developed to achieve the estimation task in models with incomplete data \citep[][]{Dempster1977}. As stated in this initial paper, the algorithm is "broadly applicable" and has since been used for mixture models estimation, around the idea that missing data also means latent non-observed (or observable) variable.\\
Given the observations $(\mathsf{y}_i)_{i \leq N}$, and an unobserved latent variable $\mathsf{\delta} \in \mathcal{D}$, the EM algorithm aims at maximizing the augmented log-likelihood
\begin{eqnarray}
\ell_N({\bm \theta})= \sum_i \log h_{\bm \theta}(\mathsf{y}_i) & \textrm{where} & 
h_{\bm \theta}(\mathsf{y}) = \int_{\mathcal{D}} f_{\bm \theta} (\mathsf{y}, \mathsf{\delta}) d\mathsf{\delta} \nonumber
\end{eqnarray}
Because $\mathsf{\delta}$ is unobserved, it is difficult to evaluate $\ell_N({\bm \theta})$. Instead, the EM maximize the expected (over $\delta$) complete-data likelihood given $\mathsf{y}$ and a previously calculated $\theta$.
\begin{eqnarray}
\label{eq:exp}
\bm{Q}({\bm \theta}; {\bm \theta'}) &=& \mathsf{E}_{\delta} \big[ \log p_{\theta}(\mathsf{y}, \delta) | {\bm \theta'}, \mathsf{y} \big] \nonumber\\
{}&=& \int_{\mathcal{D}} \log p_{\theta}(\mathsf{y}, \delta) p(\delta | {\bm \theta'}, \mathsf{y}) d\delta
\end{eqnarray}
which can usually be approximated by 
\begin{eqnarray}
\label{eq:emp}
\bm{Q}_N({\bm \theta}; {\bm \theta'}) &=& \frac{1}{N} \sum_i \int_\mathcal{D} p_{\bm \theta}(\mathsf{\delta}_i \vert \mathsf{y_i}) \log p_{\bm \theta}(\mathsf{y}_i, \mathsf{\delta}_i) d\mathsf{\delta}_i
\end{eqnarray}
because of the independence assumption between the observations. This simplification step is important because the integral becomes one-dimensional, allowing for easier calculation of its value. In order to achieve the maximization, the EM proceeds in two steps during each iteration,
\begin{equation}
\left|
\begin{array}{lcl}
\textrm{\bf(E)} & \textrm{Compute} & p_{\bm \theta}(\mathsf{\delta} \vert \mathsf{y_i}) \\
\textrm{\bf (M)} & \textrm{Set} & {\bm \theta}^{(c+1)} = \underset{\bm \theta}{\operatorname{argmax}} \quad \bm{Q}_N({\bm \theta}, {\bm \theta}^{(c)})
\end{array}
\right.
\end{equation}
We refer to \cite{Bilmes1998} for a "gentle tutorial" of the EM algorithm.\\
If we choose Model A (see Figure \ref{fig:sch_model}), using covariance matrix $\Sigma_\eta$, we cannot "jump" from Eq. \ref{eq:exp} to Eq. \ref{eq:emp} for two reasons: first, because the complete observations are not independent, even conditionally; second, because the latent variable is not straightforward to define. We have a hierarchical set of latent variables that need to be considered: $\beta$ is the higher level one; for $Z$ we want to estimate the land-use type. A technical solution would be to use Monte-Carlo EM, where the latent state variables are sampled. Although technically valid, this technique may require a huge number of samples because of the potential dimension of the latent state.\\
Another important problem is that $\mathcal{D}$ can be huge, depending on the number of land-use predictors and sites. For instance, with $p$ land-use types and $n$ sites, the number of element to sum up in Eq. \ref{eq:exp} is $p^n$ (instead of $p \times n$ with the classical EM). It is also common knowledge that the EM algorithm relies heavily on its initialization to achieve the global convergence. Poor initialization can lead to a local maxima only, even more when the dimension of the parameter space increases \citep[see][]{Wu1983a, Archambeau2003, Naim2012}. Different strategies have been proposed to overcome this problem, \citep[][]{Biernacki2001, Dicintio2012, Baudry2015}, however in the end it is most important to make the initialization as close as possible to the true solution.\\
On the other hand, by choosing Model B, and assuming that the spatial structure is just as well defined in the land-use space, we can use the Neighborhood EM algorithm, first introduced by \cite{Ambroise1997, Ambroise1998}. This algorithm allows to introduce a constraint on the mixture variable in order to account for spatial consistency. \\
In order to take the spatial dependence of objects into account, they suggest considering partitions which are optimal according to a penalized Hathaway criterion. The term of penalization should favour homogeneous classes. Spatial relationships can be summarized in different ways. In their articles, neighborhood was favored through a boolean matrix, 
\begin{equation}
v_{ij} = \left\{
\begin{array}{ll}
 1 & \textrm{if i and j are neighbours}\\
 0 & \textrm{otherwise}
\end{array}
\right.
\end{equation}
Then, defining $c_{ik} = \frac{p(\delta_i = k) f(y_i | \theta_k)}{f(y_i | \theta_k)}$ $\bm v$ is turned into a penalized term, 
\begin{eqnarray}
G(c) &=& \sum_i \sum_j \sum_k c_{ik} c_{jk} v_{ij}
\end{eqnarray}
and added to $Q$ leading to the following functional to optimize,
\begin{eqnarray}
U({\bm \theta}, {\bm \theta}^{(c)}) = Q({\bm \theta}, {\bm \theta}^{(c)}) + \beta G(\bm c) 
\end{eqnarray}
The overall principle remains the same, with two steps, one for expectation, one for maximisation. \cite{Ambroise1998} proposed an optimisation method based on the fixed point approach to achieve the E-step, which is slightly changed because of the penalization term.
\subsection{The Bayesian hierarchical model}
In the Bayesian framework, latent state models are referred to as Bayesian hierarchical models. In their general formulation, they display three levels:
\begin{eqnarray}
{\bm \pi}(\theta_1, \theta_2 | {\sf y}) &\propto & \underbrace{{\bm \pi}({\sf y} | \theta_1)}_{data} \underbrace{{\bm \pi}(\theta_1 | \theta_2)}_{process} \underbrace{{\bm \pi}( \theta_2)}_{prior} 
\end{eqnarray}
The data level refers to the likelihood of the observations given the parameters at the process level. The process level refers to the latent process captured by a spatio-temporal model for the data level parameters. In our context, what we gained with this approach is the latent spatial model, needed because the available covariates (land-use) are inadequate to capture the effects \citep[see][for extreme precipitation example]{Cooley2010, Cooley2013, Lehmann2013}.
\\
From the modelling section, we remember the special form of our modelling equation. The main issue we are trying to solve lies in the right hand side of Eq. \ref{eq:spt_model_mixt}: $z(s)$ is an indicator latent variable, where levels are the different land-use types. In this process, we make the following assumptions:
\begin{itemize}
\item only one land-use can have an influence on one station;
\item the influence of the land-use is independent of the area of the land-use (although this assumption is not really mandatory, it makes the model simpler. Otherwise, an offset variable can be used to account for the influence of the area of the land);
%\item The land-use types are not spatially correlated. This is an important assumption, because it helps keeping the model identifiable, together with the following bullet point;
%\item The spatial covariance is bounded and its upper bound is assumed known. Moreover, its bound should be below the the maximum total covariance value. 
\end{itemize}
It is important to note that the first assumption is different to the assumption that only one land-use type affects a catchment. If multiple stations are located in one catchment, many land-use effects can be observed and inferred using this model.\\
The spatial dependency structure of $Z$ is given by modelling it as a hidden Potts model, and in particular as a Gibbs random field. There is  a wealth of literature in applied statistics on these methods \citep[see][]{Cressie1993, Rue2005, Green2002}. In a Gibbs model, the probability density function of Z can be written:
\begin{eqnarray}
f(z) &=& \frac{1}{Z} \exp\{ -\sum_{c \in \mathcal{C}} U_c(z) \}
\end{eqnarray}
where $\mathcal{C}$ is the neighbourhood of $c$, and $U$ is a potential function. Here, we define, 
\begin{eqnarray}
f(z | \delta) &=& \frac{1}{Z_\delta} \exp\{ \delta^T S(z) \}
\end{eqnarray}
where $S(z) = \sum_{c' \in \mathcal{C}} 1_{z_c = z_{c'}}$ is the number  of neighbours of $c$ that belong to the same mixture.
The posterior distribution is defined over the parameters $\theta = \{v_k, \beta_k, \delta \}$. Outputs from the model also include the latent variable $z$ and the temporal basis function $f_j(k,.)$ . A very convenient way to sample from the posterior distribution is to use a Gibbs sampler. In our model, we assume $\Sigma_{\nu, k} = \nu_k Id$, and we have a Gibbs sampler that is almost explicit,
\begin{itemize}
\item $\nu_k \sim IG(\frac{N}{2} + a, \frac{\sum (y_i - \mu_{z_{i,k},k})^2}{2} + b)$, where $\mu_{z_{i,k},k}$ is the predicted value for $y_i$ using the estimated parameters;
\item $\mu_{z_{i,k},k} \sim N(\bar{y}_{j,k}, \nu_k / n_{j,k})$ and $\beta_k$ through a linear transformation of $\mu_{z_{i,k},k}$;
\item $\delta$ is updated using the scheme of \cite{Murray2006};
\item $z_{i,k} \sim \mathcal{M}(1, w_{i,1,k}, \dots,w_{i,k,k} )$ with
\begin{eqnarray}
w_{i,j,k} &=& \frac{\exp [ -\frac{1}{2}(\frac{y_i - \mu_{j,k}}{\nu_k})^2  + \delta \sum_{c \in \mathcal{C}_i} 1_{z_{c,k} = j} ]}{\sum_j \exp [ -\frac{1}{2}(\frac{y_i - \mu_{j,k}}{\nu_k})^2  + \delta \sum_{c \in \mathcal{C}_i} 1_{z_{c,k} = j} ]}.
\end{eqnarray}
\end{itemize}
The main difficulty lies in updating $\delta$. However, quick mixing can be achieved using the scheme of \cite{Murray2006} and the Swendsen-Wang algorithm to simulate from the Potts model. Additional details can be found in \cite{Murray2006, Barbu2007, Everitt2012a, Cucala2012}.
\section{Results}
\label{sec:results}
The results of three models are presented in Table \ref{tble:map}: one mixture model, one mixture model with spatial dependency and the CLR model. In the table, we present the maximum a posteriori (MAP) estimates for the land-use types, for the different temporal basis function: in our model, three basis function are used, constant, trend and seasonal; in the CLR model two temporal basis functions are used only, to match the degrees of freedom of the models. On the bottom side of the table we display the sum of square errors (SSE, using the MAP) for the prediction power of each model. The three presented models adjust differently to the data. The mixtures approaches have a better fit than the CLR model. Results show that the mixture and the spatial mixture provide almost identical results. This suggests that the estimation of the latent variables demonstrates a natural spatial correlation, which didn't need to be enforced through the spatial modelling. This spatial consistency is also demonstrated in Figure \ref{fig:pi}.
\begin{table}[!htb]
\small
 \centering 
  \caption{MAP estimates for the three fitted models. It has been ordered from the lowest baseline value, to the highest, following the mixture models estimates. We observe that the order and the amplitude of the estimates is not the same between the mixture models and the compositional-log-ratio (CLR) transform model.} 
  \label{tble:map} 
\begin{tabularx}{\textwidth}{Xccc}
\\[-1.8ex]\hline 
\hline 
 & \multicolumn{3}{c}{\textit{Dependent variable: Nitrate concentration}} \\ 
\cline{2-4}
& \textit{Spatial mixture} & \textit{Mixture model} &  \textit{CLR transform}\\
& \emph{model} & & \emph{model}\\ 
\hline 
        Managed resource protection & -4.35  & -4.35  & -0.73  \\ 
        Nature Conservation & -3.38 & -3.38 & -0.75\\
        Grazing modified pastures & -3.10  & -3.11  &  -0.49  \\ 
        Plantation forestry & -1.86  & -1.89  &  - \\
        Services, Transport and Comm. & -1.54 & -1.50 & -0.09   \\ 
        Residential & -0.98  & -0.98  &  -0.22  \\ 
        Irrigated perennial horticulture & 0.47  & 0.47  & 0.02 \\ 
  \hline
SSE & 104.62 & 103.91 & 121.81\\ 
\hline 
\hline \\[-1.8ex] 
\end{tabularx} 
\end{table} 
\subsection{Estimated land-use influence on the water quality}
\begin{figure}[htb!]
\centering
\noindent\makebox[\textwidth]{\includegraphics[width = \textwidth]{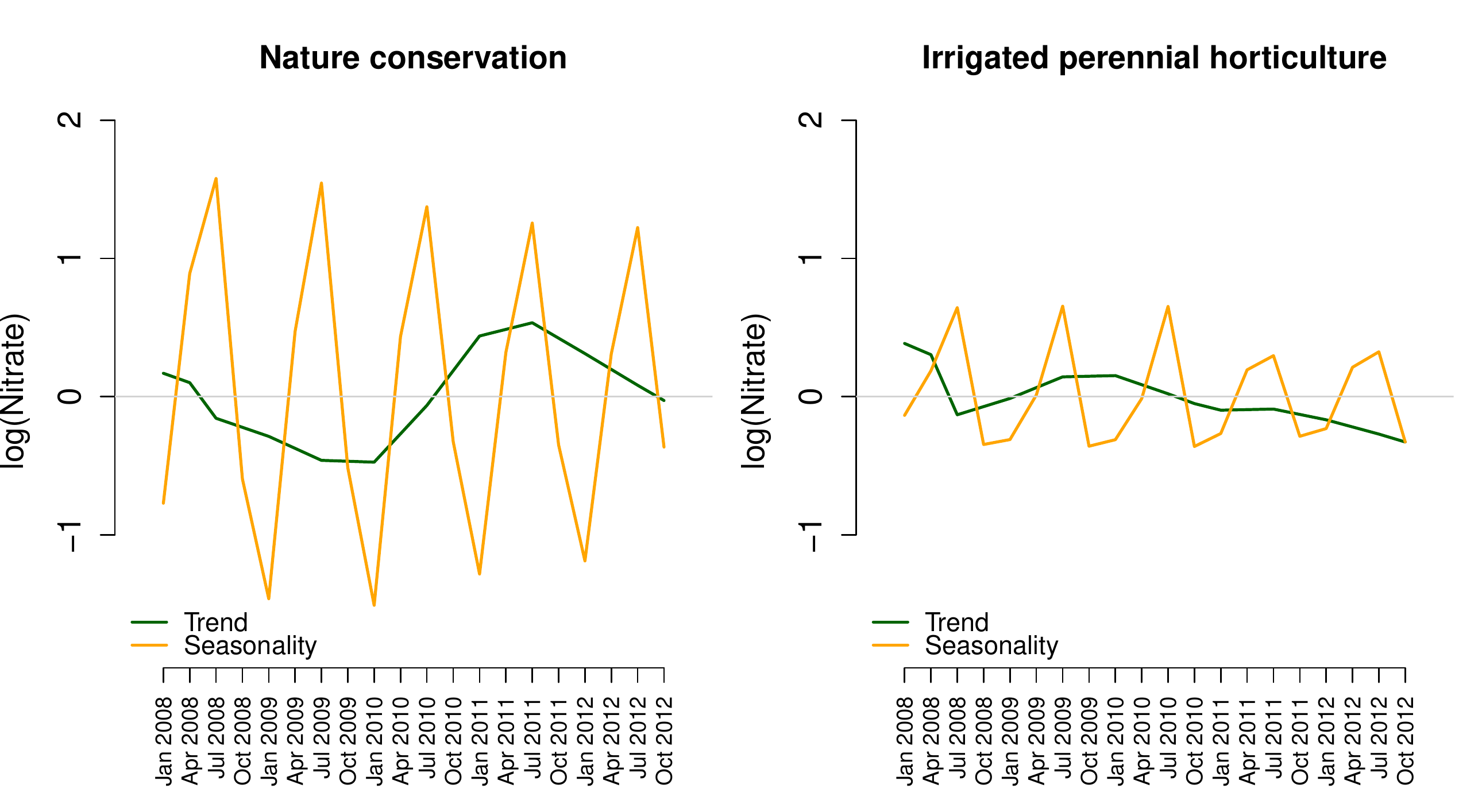}}
\caption{\label{fig:tempBasis} Example of plot of the (normalized) temporal basis functions for the land-uses "nature conservation" and "irrigated perennial horticulture". The orange line represents the seasonality, the green line represents the trend.}
\end{figure}

There are two main outputs from the model that can be analysed in order to understand the influence of land-use on the water quality. The baseline values and the temporal basis functions provide different indications for the ecological expert. The baseline value indicates a de-trended, de-seasonalised level of concentration, which can be used as a summary of the sites water quality. The temporal basis functions indicate the temporal variations, both in trend and seasonality. 
From Table \ref{tble:map}, we observe a strong consistency between the mixture models, while the MAPs estimates for the CLR transform model demonstrate less amplitude and a different order \footnote{for mathematical constraint reasons the Plantation forestry estimate had to be ignored - All the values are measured and computed on the log-scale.}. 
Figure \ref{fig:tempBasis} shows an example of estimated trend and seasonality for the land-uses "nature conservation" and "irrigated perennial horticulture". These figures indicate that:
\begin{itemize}
\item "Nature conservation" has a bigger amplitude in seasonality than "irrigated perennial horticulture";
\item The amplitude for both land-uses appears to be decreasing;
\item Their seasonality appears to have the same phase for both land-uses;
\item "Nature conservation" has a trend that indicates a longer seasonality. This is potentially due to external factors and requires further investigation;
\item "Irrigated perennial horticulture" shows a decreasing trend.
\end{itemize}
The different interpretations of the temporal basis function plots can be linked to conditions in the field in order to provide more insight into what is happening (or will happen) to water quality in this area. This also can allow for the identification of change points (observable through the trend plot) and for the prediction of water quality in changing conditions.
\subsection{Measure of uncertainty}
In Eq. \ref{eq:spt_model_mixt}, the random variable $\nu$ is representing the uncertainty of the observations. This is usually linked to unknown predictors and measurement errors. However, when the model is over-fitting the data, the uncertainty can be under estimated. This situation is encountered by the CLR model when an additional temporal basis function is added. The resulting SSE and $\sigma_{\nu}^2$ are both estimated to be equal to $0$. With the proposed model the over-fitting risk can be monitored by looking at the smoothness of the seasonality and trend curves. Figure \ref{fig:tempBasis} shows regular curves, a result consistently overruling the over-fitting risk. 
An additional feature of our model is interesting for measuring uncertainty. The mixture approach allows us to decided whether a single measurement error is modelled, or if each component of the mixture has a different measurement error. This additional flexibility allows us to identify sites / land-uses where additional sampling effort are required in order to improve the monitoring.
\subsection{Predicting the land-use type location}
\begin{figure}[htb!]
\centering
\noindent\makebox[\textwidth]{\includegraphics[width = \textwidth, page = 2]{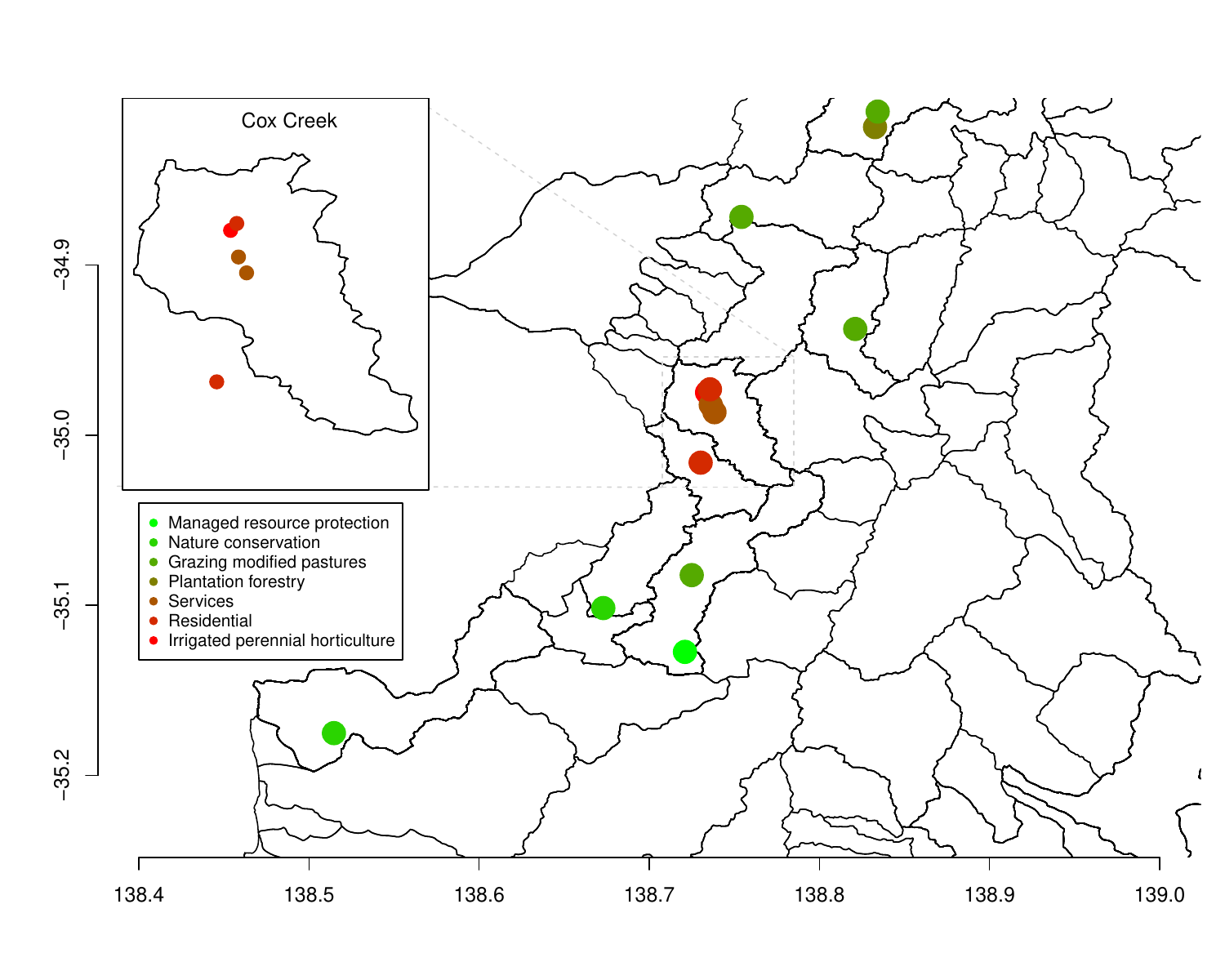}}
\caption{\label{fig:pi} Map of some of the sub-catchment in the Mount Lofty region. In the legend, items have been ordered according to Nitrate concentration: lower Nitrate concentration values are shaded green (eg MRP); higher Nitrate concentration are shaded red (eg IPH). Higher nitrate measurements are observed near the Cox Creek sub-catchment, with land-use being related to human activities (residential, services, transport and communication, irrigated perennial horticulture).}
\end{figure}

Another interesting outcome of the model is the map of estimated latent land-use type. For each station, the model can predict the most likely land-use type. Because of the nature of the model, the latent state is estimated for each Voronoi cell in each sub-catchment, allowing us to display a map of most likely land-use types. We display in Figure \ref{fig:pi} a map of the sub-catchments, coloured by their estimated latent state and ordered by their baseline values. We notice the strong spatial consistency of the map, with the higher concentration of nitrate located in the same area.
\section{Discussion}
% What we presented
In this article, we presented a model that takes into account the land-use influence on the water quality parameters when the supporting data is limited and misaligned. We presented a mixture model approach that allows the latent variables to be spatially correlated. A Bayesian algorithm was proposed to fit the model, and the method was applied to real data.
% Why it is good
We believe this model delivers two important messages. First, compositional data do not imply compositional modelling, even if the latter is easily implementable. Spatial misalignment is an important issue that should be considered carefully. Additionally, the presented model provides a helpful insight on the land-use influence when the information about it is limited, particularly in precision. The outputs of the model provide an idea about the land-use and sites association, and associates the variation of the measurements to the most likely land-use influence. Being able to identify these is of prime importance in particular for understanding the real impact of land-use on water quality.
% Drawbacks
Some improvements are possible. First, we limit the number of influence per site to one, as our aim is to identify the main land-use influence per site. This may prove a bit restrictive as some sites may be exposed to more than one significant land-use influence. A simple example for that is when the water source is a river, and the two opposite banks have two different land-use types. Our approach can be generalized to more than one influence, however this would increase the computational load significantly, even more so if we consider that different sites may have a different number of influences (and hence adding to the burden the need to estimate the number of influences for each site). The latent variable approach forces the Markov chain to run longer, primarily due to the Metropolis step within the Gibbs sampler. It is important to monitor the behaviour of the samples of the posterior distribution in order to ensure convergence of the chain. Another potential improvement would be the use of an automatic selection tool for the latent space dimension. In this article, we choose  which and how many land-use type constitute the latent space. This approach, based on expert knowledge, can be completed by a variable selection approach (of the type LASSO - \cite{Raftery2004, Stadler2010}, or slope heuristic - \cite{Baudry2012}).
\subsection*{Acknowledgements}
This study was supported by funding from the Goyder Institute for Water Research for the project "Mt Lofty Ranges Water Allocation and Planning" led by Associate Professor Jim Cox. The authors would like to thank SA Water, SA EPA, DEWNR, AMLR NRMB and Dr Leon van der Linden for providing water quality data from the Mt Lofty Ranges for the risk assessment. Thank you also to staff from numerous South Australian Government Agencies who provided valuable feedback at presentations given about project outputs. 
% BibTeX users please use one of
   % basic style, author-year citations
\bibliographystyle{model2-names}
%\bibliography{/home/ick003/Documents/Donnees/References/Publications-Compositional_Data_Analysis}
\bibliography{Comp_data_v7_arXiv.bbl}

\begin{thebibliography}{56}
\expandafter\ifx\csname natexlab\endcsname\relax\def\natexlab#1{#1}\fi
\providecommand{\url}[1]{\texttt{#1}}
\providecommand{\href}[2]{#2}
\providecommand{\path}[1]{#1}
\providecommand{\DOIprefix}{doi:}
\providecommand{\ArXivprefix}{arXiv:}
\providecommand{\URLprefix}{URL: }
\providecommand{\Pubmedprefix}{pmid:}
\providecommand{\doi}[1]{\href{http://dx.doi.org/#1}{\path{#1}}}
\providecommand{\Pubmed}[1]{\href{pmid:#1}{\path{#1}}}
\providecommand{\bibinfo}[2]{#2}
\ifx\xfnm\relax \def\xfnm[#1]{\unskip,\space#1}\fi
%Type = Article
\bibitem[{Aitchison(2003)}]{Aitchison2003}
\bibinfo{author}{Aitchison, J.}, \bibinfo{year}{2003}.
\newblock \bibinfo{title}{{A concise guide to compositional data analysis}}.
\newblock \bibinfo{journal}{CDA Workshop, Girona} .
%Type = Inproceedings
\bibitem[{Ambroise et~al.(1997)Ambroise, Dang and Govaert}]{Ambroise1997}
\bibinfo{author}{Ambroise, C.}, \bibinfo{author}{Dang, V.M.},
  \bibinfo{author}{Govaert, G.}, \bibinfo{year}{1997}.
\newblock \bibinfo{title}{{Clustering of spatial data by the EM algorithm}},
  in: \bibinfo{booktitle}{geoENV I - Geostatistics for Environmental
  Applications, Quantitative Geology and Geostatistics (Vol. 9)}, pp.
  \bibinfo{pages}{493--504}.
%Type = Article
\bibitem[{Ambroise and Govaert(1998)}]{Ambroise1998}
\bibinfo{author}{Ambroise, C.}, \bibinfo{author}{Govaert, G.},
  \bibinfo{year}{1998}.
\newblock \bibinfo{title}{{Convergence of an EM-type algorithm for spatial
  clustering}}.
\newblock \bibinfo{journal}{Pattern Recognition Letters} \bibinfo{volume}{19},
  \bibinfo{pages}{919--927}.
%Type = Inproceedings
\bibitem[{Archambeau et~al.(2003)Archambeau, Lee and
  Verleysen}]{Archambeau2003}
\bibinfo{author}{Archambeau, C.}, \bibinfo{author}{Lee, J.A.},
  \bibinfo{author}{Verleysen, M.}, \bibinfo{year}{2003}.
\newblock \bibinfo{title}{{On Convergence Problems with the EM for Finite
  Gaussian Mixtures}}, in: \bibinfo{booktitle}{Neural Networks}, pp.
  \bibinfo{pages}{99--106}.
%Type = Article
\bibitem[{Arnold et~al.(1995)Arnold, Williams and Maidment}]{Arnold1995}
\bibinfo{author}{Arnold, J.G.}, \bibinfo{author}{Williams, J.R.},
  \bibinfo{author}{Maidment, D.R.}, \bibinfo{year}{1995}.
\newblock \bibinfo{title}{{Continuous-Time Water and Sediment-Routing Model for
  Large Basins}}.
\newblock \bibinfo{journal}{Journal of Hydraulic Engineering}
  \bibinfo{volume}{121}, \bibinfo{pages}{171--183}.
%Type = Article
\bibitem[{Bakar and Sahu(2013)}]{Bakar2013}
\bibinfo{author}{Bakar, K.}, \bibinfo{author}{Sahu, S.}, \bibinfo{year}{2013}.
\newblock \bibinfo{title}{{spTimer: Spatio-Temporal Bayesian Modelling Using
  R}}.
\newblock \bibinfo{journal}{Journal of Statistical Software} .
%Type = Article
\bibitem[{Barbu and Zhu(2007)}]{Barbu2007}
\bibinfo{author}{Barbu, A.}, \bibinfo{author}{Zhu, S.C.}, \bibinfo{year}{2007}.
\newblock \bibinfo{title}{{Generalizing Swendsen-Wang for Image Analysis}}.
\newblock \bibinfo{journal}{Journal of Computational and Graphical Statistics}
  \bibinfo{volume}{16}, \bibinfo{pages}{877--900}.
%Type = Techreport
\bibitem[{Baudry and Celeux(2015)}]{Baudry2015}
\bibinfo{author}{Baudry, J.P.}, \bibinfo{author}{Celeux, G.},
  \bibinfo{year}{2015}.
\newblock \bibinfo{title}{{EM for mixtures - Initialization requires special
  care}}.
\newblock \bibinfo{type}{Technical Report}. INRIA.
%Type = Article
\bibitem[{Baudry et~al.(2012)Baudry, Maugis and Michel}]{Baudry2012}
\bibinfo{author}{Baudry, J.P.}, \bibinfo{author}{Maugis, C.},
  \bibinfo{author}{Michel, B.}, \bibinfo{year}{2012}.
\newblock \bibinfo{title}{{Slope heuristics: Overview and implementation}}.
\newblock \bibinfo{journal}{Statistics and Computing} \bibinfo{volume}{22},
  \bibinfo{pages}{455--470}.
%Type = Article
\bibitem[{Beck(1987)}]{Beck1987}
\bibinfo{author}{Beck, M.B.}, \bibinfo{year}{1987}.
\newblock \bibinfo{title}{{Water quality modeling: A review of the analysis of
  uncertainty}}.
\newblock \bibinfo{journal}{Water Resources Research} \bibinfo{volume}{23},
  \bibinfo{pages}{1393--1442}.
%Type = Techreport
\bibitem[{Biernacki et~al.(2001)Biernacki, Celeux and Govaert}]{Biernacki2001}
\bibinfo{author}{Biernacki, C.}, \bibinfo{author}{Celeux, G.},
  \bibinfo{author}{Govaert, G.}, \bibinfo{year}{2001}.
\newblock \bibinfo{title}{{Strategies for Getting the Highest Likelihood in
  Mixture Models}}.
\newblock \bibinfo{type}{Technical Report} \bibinfo{number}{RR-4255}. INRIA.
%Type = Article
\bibitem[{Bilmes(1998)}]{Bilmes1998}
\bibinfo{author}{Bilmes, J.A.}, \bibinfo{year}{1998}.
\newblock \bibinfo{title}{{A gentle tutorial of the EM algorithm and its
  application to parameter estimation for Gaussian mixture and hidden Markov
  models}}.
\newblock \bibinfo{journal}{International Computer Science Institute}
  \bibinfo{volume}{4}, \bibinfo{pages}{1--15}.
%Type = Article
\bibitem[{Bonyah et~al.(2013)Bonyah, Munyakazi, Asong and Bashiru}]{Bonyah2013}
\bibinfo{author}{Bonyah, E.}, \bibinfo{author}{Munyakazi, L.},
  \bibinfo{author}{Asong, D.}, \bibinfo{author}{Bashiru, I.I.},
  \bibinfo{year}{2013}.
\newblock \bibinfo{title}{{Application of area to point Kriging to breast
  cancer incidence in Ashanti Region of Ghana}}.
\newblock \bibinfo{journal}{International Journal of Medicine and Medical
  Sciences} \bibinfo{volume}{5}, \bibinfo{pages}{67--74}.
%Type = Book
\bibitem[{van~den Boogaart and Tolosana-Delgado(2013)}]{VandenBoogaart2013}
\bibinfo{author}{van~den Boogaart, K.G.}, \bibinfo{author}{Tolosana-Delgado,
  R.}, \bibinfo{year}{2013}.
\newblock \bibinfo{title}{{Analyzing Compositional Data with R}}.
\newblock \bibinfo{publisher}{Springer Berlin Heidelberg},
  \bibinfo{address}{Berlin, Heidelberg}.
%Type = Article
\bibitem[{Buck et~al.(2004)Buck, Niyogi and Townsend}]{Buck2004}
\bibinfo{author}{Buck, O.}, \bibinfo{author}{Niyogi, D.K.},
  \bibinfo{author}{Townsend, C.R.}, \bibinfo{year}{2004}.
\newblock \bibinfo{title}{{Scale-dependence of land use effects on water
  quality of streams in agricultural catchments.}}
\newblock \bibinfo{journal}{Environmental pollution (Barking, Essex : 1987)}
  \bibinfo{volume}{130}, \bibinfo{pages}{287--99}.
%Type = Article
\bibitem[{Burcher(2009)}]{Burcher2009}
\bibinfo{author}{Burcher, C.L.}, \bibinfo{year}{2009}.
\newblock \bibinfo{title}{{Using simplified watershed hydrology to define
  spatially explicit 'zones of influence'}}.
\newblock \bibinfo{journal}{Hydrobiologia} \bibinfo{volume}{618},
  \bibinfo{pages}{149--160}.
%Type = Article
\bibitem[{Cleveland and Cleveland(1990)}]{Cleveland1990}
\bibinfo{author}{Cleveland, R.}, \bibinfo{author}{Cleveland, W.},
  \bibinfo{year}{1990}.
\newblock \bibinfo{title}{{STL: A seasonal-trend decomposition procedure based
  on loess}}.
\newblock \bibinfo{journal}{Journal of Official Statistics}
  \bibinfo{volume}{6}, \bibinfo{pages}{3--73}.
%Type = Techreport
\bibitem[{Cooley(2013)}]{Cooley2013}
\bibinfo{author}{Cooley, D.}, \bibinfo{year}{2013}.
\newblock \bibinfo{title}{{Modeling Both Climate and Weather Spatial Effects
  for Extreme Precipitation}}.
\newblock \bibinfo{type}{Technical Report}. Colorade State University.
%Type = Article
\bibitem[{Cooley and Sain(2010)}]{Cooley2010}
\bibinfo{author}{Cooley, D.}, \bibinfo{author}{Sain, S.R.},
  \bibinfo{year}{2010}.
\newblock \bibinfo{title}{{Spatial hierarchical modeling of precipitation
  extremes from a regional climate model}}.
\newblock \bibinfo{journal}{Journal of Agricultural, Biological, and
  Environmental Statistics} \bibinfo{volume}{15}, \bibinfo{pages}{381--402}.
%Type = Article
\bibitem[{Cox et~al.(2012)Cox, Oliver, Fleming and Anderson}]{Cox2012}
\bibinfo{author}{Cox, J.W.}, \bibinfo{author}{Oliver, D.P.},
  \bibinfo{author}{Fleming, N.K.}, \bibinfo{author}{Anderson, J.S.},
  \bibinfo{year}{2012}.
\newblock \bibinfo{title}{{Off-site transport of nutrients and sediment from
  three main land-uses in the Mt Lofty Ranges, South Australia}}.
\newblock \bibinfo{journal}{Agricultural Water Management}
  \bibinfo{volume}{106}, \bibinfo{pages}{50--59}.
%Type = Book
\bibitem[{Cressie(2015)}]{Cressie1993}
\bibinfo{author}{Cressie, N.A.C.}, \bibinfo{year}{2015}.
\newblock \bibinfo{title}{{Statistics for Spatial Data}}.
\newblock \bibinfo{publisher}{Wiley}.
%Type = Book
\bibitem[{Cressie and Wikle(2011)}]{Cressie2011}
\bibinfo{author}{Cressie, N.A.C.}, \bibinfo{author}{Wikle, C.},
  \bibinfo{year}{2011}.
\newblock \bibinfo{title}{{Statistics for Spatio-Temporal Data}}.
\newblock \bibinfo{publisher}{Wiley}.
%Type = Article
\bibitem[{Cucala and Marin(2013)}]{Cucala2012}
\bibinfo{author}{Cucala, L.}, \bibinfo{author}{Marin, J.M.},
  \bibinfo{year}{2013}.
\newblock \bibinfo{title}{{Bayesian inference on a mixture model with spatial
  dependence}}.
\newblock \bibinfo{journal}{Journal Of Computational And Graphical Statistics}
  \bibinfo{volume}{22}, \bibinfo{pages}{584--597}.
%Type = Article
\bibitem[{Dempster et~al.(1977)Dempster, Laird and Rubin}]{Dempster1977}
\bibinfo{author}{Dempster, A.P.}, \bibinfo{author}{Laird, N.M.},
  \bibinfo{author}{Rubin, D.B.}, \bibinfo{year}{1977}.
\newblock \bibinfo{title}{{Maximum Likelihood from Incomplete Data via the EM
  Algorithm}}.
\newblock \bibinfo{journal}{Journal of the Royal Statistical Society. Series B
  (Methodological)} \bibinfo{volume}{39}, \bibinfo{pages}{1--38}.
%Type = Phdthesis
\bibitem[{Dicintio(2012)}]{Dicintio2012}
\bibinfo{author}{Dicintio, S.}, \bibinfo{year}{2012}.
\newblock \bibinfo{title}{{Comparing Approaches to Initializing the
  Expectation-Maximization Algorithm}}.
\newblock Ph.D. thesis. University of Guelph.
%Type = Article
\bibitem[{Everitt(2012)}]{Everitt2012a}
\bibinfo{author}{Everitt, R.G.}, \bibinfo{year}{2012}.
\newblock \bibinfo{title}{{Bayesian Parameter Estimation for Latent Markov
  Random Fields and Social Networks}}.
\newblock \bibinfo{journal}{Journal of Computational and Graphical Statistics}
  \bibinfo{volume}{8600}, \bibinfo{pages}{26}.
%Type = Techreport
\bibitem[{Ford et~al.(2015)Ford, Ickowicz, Oliver, Hayes and
  Kookana}]{Ford2015}
\bibinfo{author}{Ford, J.}, \bibinfo{author}{Ickowicz, A.},
  \bibinfo{author}{Oliver, D.}, \bibinfo{author}{Hayes, K.},
  \bibinfo{author}{Kookana, R.}, \bibinfo{year}{2015}.
\newblock \bibinfo{title}{{Integrated catchment water planning support for
  Adelaide Mount Lofty Ranges Water Allocation Planning ( GWAP Project ) Task 5
  : Tiered Water Quality Risk Assessment}}.
\newblock \bibinfo{type}{Technical Report} \bibinfo{number}{15/4}. Goyder
  Institute for Water Research. \bibinfo{address}{Adelaide}.
%Type = Article
\bibitem[{Gelfand et~al.(2001)Gelfand, Zhu and Carlin}]{Gelfand2001}
\bibinfo{author}{Gelfand, a.E.}, \bibinfo{author}{Zhu, L.},
  \bibinfo{author}{Carlin, B.P.}, \bibinfo{year}{2001}.
\newblock \bibinfo{title}{{On the change of support problem for spatio-temporal
  data.}}
\newblock \bibinfo{journal}{Biostatistics (Oxford, England)}
  \bibinfo{volume}{2}, \bibinfo{pages}{31--45}.
%Type = Article
\bibitem[{Green and Richardson(2002)}]{Green2002}
\bibinfo{author}{Green, P.J.}, \bibinfo{author}{Richardson, S.},
  \bibinfo{year}{2002}.
\newblock \bibinfo{title}{{Hidden Markov Models and Disease Mapping}}.
\newblock \bibinfo{journal}{Journal of the American Statistical Association}
  \bibinfo{volume}{97}, \bibinfo{pages}{1055--1070}.
%Type = Article
\bibitem[{Hunsaker and Levine(1995)}]{Hunsaker1995}
\bibinfo{author}{Hunsaker, C.T.}, \bibinfo{author}{Levine, D.A.},
  \bibinfo{year}{1995}.
\newblock \bibinfo{title}{{Hierarchical Approaches of Water Quality in Rivers
  Study processes are important in developing}}.
\newblock \bibinfo{journal}{Sciences-New York} \bibinfo{volume}{45},
  \bibinfo{pages}{193--203}.
%Type = Article
\bibitem[{King et~al.(2005)King, Baker, Whigham, Weller, Jordan, Kazyak and
  Hurd}]{King2005}
\bibinfo{author}{King, R.S.}, \bibinfo{author}{Baker, M.E.},
  \bibinfo{author}{Whigham, D.F.}, \bibinfo{author}{Weller, D.E.},
  \bibinfo{author}{Jordan, T.E.}, \bibinfo{author}{Kazyak, P.F.},
  \bibinfo{author}{Hurd, M.K.}, \bibinfo{year}{2005}.
\newblock \bibinfo{title}{{Spatial Considerations for Linking Watershed Land
  Cover To Ecological Indicators in Streams}}.
\newblock \bibinfo{journal}{Ecological Applications} \bibinfo{volume}{15},
  \bibinfo{pages}{137--153}.
%Type = Article
\bibitem[{Kyriakidis(2004)}]{Kyriakidis2004}
\bibinfo{author}{Kyriakidis, P.}, \bibinfo{year}{2004}.
\newblock \bibinfo{title}{{A geostatistical framework for area to point spatial
  interpolation}}.
\newblock \bibinfo{journal}{Geographical Analysis} \bibinfo{volume}{36},
  \bibinfo{pages}{259--289}.
%Type = Inproceedings
\bibitem[{Lehmann et~al.(2013)Lehmann, Phatak, Soltyk, Chia, Lau and
  Palmer}]{Lehmann2013}
\bibinfo{author}{Lehmann, E.A.}, \bibinfo{author}{Phatak, A.},
  \bibinfo{author}{Soltyk, S.}, \bibinfo{author}{Chia, J.},
  \bibinfo{author}{Lau, R.}, \bibinfo{author}{Palmer, M.},
  \bibinfo{year}{2013}.
\newblock \bibinfo{title}{{Bayesian hierarchical modelling of rainfall
  extremes}}, in: \bibinfo{booktitle}{20th International Congress on Modelling
  and Simulation, Adelaide, Australia, 1-6 December 2013}, pp.
  \bibinfo{pages}{1--6}.
%Type = Article
\bibitem[{Lindstrom et~al.(2011)Lindstrom, Szpiro, Sampson, Sheppard, Oron,
  Richards and Larson}]{Lindstrom2011}
\bibinfo{author}{Lindstrom, J.}, \bibinfo{author}{Szpiro, A.},
  \bibinfo{author}{Sampson, P.}, \bibinfo{author}{Sheppard, L.},
  \bibinfo{author}{Oron, a.}, \bibinfo{author}{Richards, M.},
  \bibinfo{author}{Larson, T.}, \bibinfo{year}{2011}.
\newblock \bibinfo{title}{{A flexible spatio-temporal model for air pollution:
  Allowing for spatio-temporal covariates}}.
\newblock \bibinfo{journal}{UW Biostatistics Working Paper Series}
  \bibinfo{volume}{370}, \bibinfo{pages}{1--38}.
%Type = Article
\bibitem[{Lindstrom et~al.(2013)Lindstrom, Szpiro, Sampson, Bergen and
  Sheppard}]{Lindstrom2013}
\bibinfo{author}{Lindstrom, J.}, \bibinfo{author}{Szpiro, A.},
  \bibinfo{author}{Sampson, P.D.}, \bibinfo{author}{Bergen, S.},
  \bibinfo{author}{Sheppard, L.}, \bibinfo{year}{2013}.
\newblock \bibinfo{title}{{SpatioTemporal : An R Package for Spatio-Temporal
  Modelling of Air-Pollution}}.
\newblock \bibinfo{journal}{CRAN Vignettes} .
%Type = Book
\bibitem[{McLachlan and Peel(2000)}]{McLachlan2000}
\bibinfo{author}{McLachlan, G.}, \bibinfo{author}{Peel, D.},
  \bibinfo{year}{2000}.
\newblock \bibinfo{title}{{Finite Mixture Models}}.
\newblock \bibinfo{publisher}{John Wiley {\&} Sons}, \bibinfo{address}{New
  York}.
%Type = Article
\bibitem[{Meinicke et~al.(2011)Meinicke, A{\ss}hauer and
  Lingner}]{Meinicke2011}
\bibinfo{author}{Meinicke, P.}, \bibinfo{author}{A{\ss}hauer, K.P.},
  \bibinfo{author}{Lingner, T.}, \bibinfo{year}{2011}.
\newblock \bibinfo{title}{{Mixture models for analysis of the taxonomic
  composition of metagenomes}}.
\newblock \bibinfo{journal}{Bioinformatics} \bibinfo{volume}{27},
  \bibinfo{pages}{1618--1624}.
%Type = Article
\bibitem[{Murray et~al.(2006)Murray, Ghahramani and MacKay}]{Murray2006}
\bibinfo{author}{Murray, I.}, \bibinfo{author}{Ghahramani, Z.},
  \bibinfo{author}{MacKay, D.J.C.}, \bibinfo{year}{2006}.
\newblock \bibinfo{title}{{MCMC for doubly-intractable distributions}}.
\newblock \bibinfo{journal}{Proceedings of the 22nd Annual Conference on
  Uncertainty in Artificial Intelligence (UAI-06)} , \bibinfo{pages}{359--366}.
%Type = Article
\bibitem[{Naim and Gildea(2012)}]{Naim2012}
\bibinfo{author}{Naim, I.}, \bibinfo{author}{Gildea, D.}, \bibinfo{year}{2012}.
\newblock \bibinfo{title}{{Convergence of the EM Algorithm for Gaussian
  Mixtures with Unbalanced Mixing Coefficients}}.
\newblock \bibinfo{journal}{Proceedings of the 29th International Conference on
  Machine Learning (ICML-12)} , \bibinfo{pages}{1655--1662}.
%Type = Article
\bibitem[{Neelon et~al.(2014)Neelon, Gelfand and Miranda}]{Neelon2014}
\bibinfo{author}{Neelon, B.}, \bibinfo{author}{Gelfand, A.E.},
  \bibinfo{author}{Miranda, M.L.}, \bibinfo{year}{2014}.
\newblock \bibinfo{title}{{A multivariate spatial mixture model for areal data:
  Examining regional differences in standardized test scores}}.
\newblock \bibinfo{journal}{Journal of the Royal Statistical Society. Series C:
  Applied Statistics} \bibinfo{volume}{63}, \bibinfo{pages}{737--761}.
%Type = Article
\bibitem[{Ongaro et~al.(2008)Ongaro, Migliorati and Monti}]{Ongaro2008}
\bibinfo{author}{Ongaro, A.}, \bibinfo{author}{Migliorati, S.},
  \bibinfo{author}{Monti, G.S.}, \bibinfo{year}{2008}.
\newblock \bibinfo{title}{{A new distribution on the simplex containing the
  Dirichlet family}}.
\newblock \bibinfo{journal}{CoDaWork 2008, the 3rd International Workshop on
  Compositional Data Analysis} .
%Type = Article
\bibitem[{Peterson et~al.(2011)Peterson, Sheldon, Darnell, Bunn and
  Harch}]{Peterson2011}
\bibinfo{author}{Peterson, E.E.}, \bibinfo{author}{Sheldon, F.},
  \bibinfo{author}{Darnell, R.}, \bibinfo{author}{Bunn, S.E.},
  \bibinfo{author}{Harch, B.D.}, \bibinfo{year}{2011}.
\newblock \bibinfo{title}{{A comparison of spatially explicit landscape
  representation methods and their relationship to stream condition}}.
\newblock \bibinfo{journal}{Freshwater Biology} \bibinfo{volume}{56},
  \bibinfo{pages}{590--610}.
%Type = Article
\bibitem[{Peterson et~al.(2007)Peterson, Theobald and {Ver
  Hoef}}]{Peterson2007}
\bibinfo{author}{Peterson, E.E.}, \bibinfo{author}{Theobald, D.M.},
  \bibinfo{author}{{Ver Hoef}, J.M.}, \bibinfo{year}{2007}.
\newblock \bibinfo{title}{{Geostatistical modelling on stream networks:
  Developing valid covariance matrices based on hydrologic distance and stream
  flow}}.
\newblock \bibinfo{journal}{Freshwater Biology} \bibinfo{volume}{52},
  \bibinfo{pages}{267--279}.
%Type = Article
\bibitem[{Raftery and Dean(2004)}]{Raftery2004}
\bibinfo{author}{Raftery, A.E.}, \bibinfo{author}{Dean, N.},
  \bibinfo{year}{2004}.
\newblock \bibinfo{title}{{Variable Selection for Model-Based Clustering}}.
\newblock \bibinfo{journal}{Journal of the American Statistical Association}
  \bibinfo{volume}{101}, \bibinfo{pages}{168--178}.
%Type = Article
\bibitem[{Rue(2005)}]{Rue2005}
\bibinfo{author}{Rue, H.}, \bibinfo{year}{2005}.
\newblock \bibinfo{title}{{Gaussian Markov Random Fields: Theory and
  Applications}}.
\newblock \bibinfo{journal}{Hand The} \bibinfo{volume}{104},
  \bibinfo{pages}{263 p.}
%Type = Article
\bibitem[{Sampson et~al.(2011)Sampson, Szpiro, Sheppard, Lindstr{\"{o}}m and
  Kaufman}]{Sampson2011}
\bibinfo{author}{Sampson, P.D.}, \bibinfo{author}{Szpiro, A.a.},
  \bibinfo{author}{Sheppard, L.}, \bibinfo{author}{Lindstr{\"{o}}m, J.},
  \bibinfo{author}{Kaufman, J.D.}, \bibinfo{year}{2011}.
\newblock \bibinfo{title}{{Pragmatic estimation of a spatio-temporal air
  quality model with irregular monitoring data}}.
\newblock \bibinfo{journal}{Atmospheric Environment} \bibinfo{volume}{45},
  \bibinfo{pages}{6593--6606}.
%Type = Article
\bibitem[{Shen et~al.(2014)Shen, Hou, Li and Aini}]{Shen2014}
\bibinfo{author}{Shen, Z.}, \bibinfo{author}{Hou, X.}, \bibinfo{author}{Li,
  W.}, \bibinfo{author}{Aini, G.}, \bibinfo{year}{2014}.
\newblock \bibinfo{title}{{Relating landscape characteristics to non-point
  source pollution in a typical urbanized watershed in the municipality of
  Beijing}}.
\newblock \bibinfo{journal}{Landscape and Urban Planning}
  \bibinfo{volume}{123}, \bibinfo{pages}{96--107}.
%Type = Article
\bibitem[{St{\"{a}}dler et~al.(2010)St{\"{a}}dler, B{\"{u}}hlmann and van~de
  Geer}]{Stadler2010}
\bibinfo{author}{St{\"{a}}dler, N.}, \bibinfo{author}{B{\"{u}}hlmann, P.},
  \bibinfo{author}{van~de Geer, S.}, \bibinfo{year}{2010}.
\newblock \bibinfo{title}{{l1-Penalization for Mixture Regression Models}}.
\newblock \bibinfo{journal}{Test} \bibinfo{volume}{19},
  \bibinfo{pages}{209--256}.
%Type = Article
\bibitem[{Strayer et~al.(2003)Strayer, Beighley, Thompson, Brooks, Nilsson,
  Pinay and Naiman}]{Strayer2003}
\bibinfo{author}{Strayer, D.L.}, \bibinfo{author}{Beighley, R.E.},
  \bibinfo{author}{Thompson, L.C.}, \bibinfo{author}{Brooks, S.},
  \bibinfo{author}{Nilsson, C.}, \bibinfo{author}{Pinay, G.},
  \bibinfo{author}{Naiman, R.J.}, \bibinfo{year}{2003}.
\newblock \bibinfo{title}{{Effects of Land Cover on Stream Ecosystems: Roles of
  Empirical Models and Scaling Issues}}.
\newblock \bibinfo{journal}{Ecosystems} \bibinfo{volume}{6},
  \bibinfo{pages}{407--423}.
%Type = Article
\bibitem[{Szpiro et~al.(2010)Szpiro, Sampson, Sheppard, Lumley, Adar and
  Kaufman}]{Szpiro2010}
\bibinfo{author}{Szpiro, A.A.}, \bibinfo{author}{Sampson, P.D.},
  \bibinfo{author}{Sheppard, L.}, \bibinfo{author}{Lumley, T.},
  \bibinfo{author}{Adar, S.D.}, \bibinfo{author}{Kaufman, J.D.},
  \bibinfo{year}{2010}.
\newblock \bibinfo{title}{{Predicting intra-urban variation in air pollution
  concentrations with complex spatio-temporal dependencies}}.
\newblock \bibinfo{journal}{Environmetrics} \bibinfo{volume}{21},
  \bibinfo{pages}{606--631}.
%Type = Article
\bibitem[{Tong and Chen(2002)}]{Tong2002}
\bibinfo{author}{Tong, S.T.}, \bibinfo{author}{Chen, W.}, \bibinfo{year}{2002}.
\newblock \bibinfo{title}{{Modeling the relationship between land use and
  surface water quality}}.
\newblock \bibinfo{journal}{Journal of Environmental Management}
  \bibinfo{volume}{66}, \bibinfo{pages}{377--393}.
%Type = Article
\bibitem[{Varcoe et~al.(2010)Varcoe, van Leeuwen, Chittleborough, Cox, Smernik
  and Heitz}]{Varcoe2010}
\bibinfo{author}{Varcoe, J.}, \bibinfo{author}{van Leeuwen, J.A.},
  \bibinfo{author}{Chittleborough, D.J.}, \bibinfo{author}{Cox, J.W.},
  \bibinfo{author}{Smernik, R.J.}, \bibinfo{author}{Heitz, A.},
  \bibinfo{year}{2010}.
\newblock \bibinfo{title}{{Changes in water quality following gypsum
  application to catchment soils of the Mount Lofty Ranges, South Australia}}.
\newblock \bibinfo{journal}{Organic Geochemistry} \bibinfo{volume}{41},
  \bibinfo{pages}{116--123}.
%Type = Article
\bibitem[{Woolrich et~al.(2005)Woolrich, Behrens, Beckmann and
  Smith}]{Woolrich2005}
\bibinfo{author}{Woolrich, M.}, \bibinfo{author}{Behrens, T.},
  \bibinfo{author}{Beckmann, C.}, \bibinfo{author}{Smith, S.},
  \bibinfo{year}{2005}.
\newblock \bibinfo{title}{{Mixture models with adaptive spatial regularization
  for segmentation with an application to fMRI data}}.
\newblock \bibinfo{journal}{Medical Imaging, IEEE Transactions on}
  \bibinfo{volume}{24}, \bibinfo{pages}{1--11}.
%Type = Article
\bibitem[{Wu(1983)}]{Wu1983a}
\bibinfo{author}{Wu, C.}, \bibinfo{year}{1983}.
\newblock \bibinfo{title}{{On the convergence properties of the EM algorithm}}.
\newblock \bibinfo{journal}{Annals of Statistics} \bibinfo{volume}{11},
  \bibinfo{pages}{95--103}.
%Type = Article
\bibitem[{Yoo and Kyriakidis(2006)}]{Yoo2006}
\bibinfo{author}{Yoo, E.H.}, \bibinfo{author}{Kyriakidis, P.C.},
  \bibinfo{year}{2006}.
\newblock \bibinfo{title}{{Area-to-point Kriging with inequality-type data}}.
\newblock \bibinfo{journal}{Journal of Geographical Systems}
  \bibinfo{volume}{8}, \bibinfo{pages}{357--390}.
%Type = Article
\bibitem[{Zhu et~al.(2003)Zhu, Carlin and Gelfand}]{Zhu2003}
\bibinfo{author}{Zhu, L.}, \bibinfo{author}{Carlin, B.},
  \bibinfo{author}{Gelfand, A.}, \bibinfo{year}{2003}.
\newblock \bibinfo{title}{{Hierarchical regression with misaligned spatial
  data: relating ambient ozone and pediatric asthma ER visits in Atlanta}}.
\newblock \bibinfo{journal}{Environmetrics} , \bibinfo{pages}{1--33}.

\end{thebibliography}
\end{document}